\documentclass[prd,preprint, superscriptaddress,preprintnumbers,nofootinbib]{revtex4}
\usepackage{graphicx}
\usepackage{epsfig}
\usepackage{bm}
\usepackage{latexsym,amssymb,amsmath,amssymb,wasysym,float}

\usepackage{color}

\usepackage{enumitem}

\newcommand{\postscript}[2]{\setlength{\epsfxsize}{#2\hsize}
   \centerline{\epsfbox{#1}}}

\usepackage[usenames,dvipsnames]{xcolor}
\definecolor{orange}{cmyk}{0,0.5,1,0}
\definecolor{rossoCP3}{cmyk}{0,.88,.77,.40}
\definecolor{graa}{rgb}{0.8,0.8,0.8}
\definecolor{blaa}{rgb}{0.2,0.2,0.6}

\begin{document}

\preprint{MPP-2024-67}
\preprint{LMU-ASC 04/24}

\title{\color{rossoCP3}  More on Black Holes Perceiving the Dark Dimension}

\author{\bf Luis A. Anchordoqui}

\affiliation{Department of Physics and Astronomy,\\  Lehman College, City University of
  New York, NY 10468, USA
}

\affiliation{Department of Physics,\\
 Graduate Center,  City University of
  New York,  NY 10016, USA
}

\affiliation{Department of Astrophysics,
 American Museum of Natural History, NY
 10024, USA
}

\author{\bf Ignatios Antoniadis}

\affiliation{High Energy Physics Research Unit, Faculty of Science, Chulalongkorn University, Bangkok 1030, Thailand}

\affiliation{Laboratoire de Physique Th\'eorique et Hautes \'Energies
  - LPTHE \\
Sorbonne Universit\'e, CNRS, 4 Place Jussieu, 75005 Paris, France
}

\author{\bf Dieter\nolinebreak~L\"ust}

\affiliation{Max--Planck--Institut f\"ur Physik,  
 Werner--Heisenberg--Institut,
80805 M\"unchen, Germany
}

\affiliation{Arnold Sommerfeld Center for Theoretical Physics, 
Ludwig-Maximilians-Universit\"at M\"unchen,
80333 M\"unchen, Germany
}

\begin{abstract}
\vskip 2mm \noindent In the last two years the dark dimension scenario has emerged as focal
point of many research interests. In particular, it functions as a
stepping stone to address the cosmological hierarchy problem and
provides a colosseum for dark matter contenders. We reexamine the
possibility that primordial black holes (PBHs) perceiving
the dark dimension could constitute all of the dark matter in the
universe. We re-assess limits on the abundance of PBHs as dark matter candidates from $\gamma$-ray emission resulting
from Hawking evaporation. We re-evaluate constraints from the diffuse $\gamma$-ray
emission in the direction of the Galactic center which offer the best and
most solid upper limits on the dark matter fraction composed of
PBHs. The revised mass range which allows PBHs to assemble all cosmological dark matter is
estimated to be $10^{15} \alt M_{\rm BH}/{\rm g} \alt 10^{21}$. We
demonstrate that due to the constraints from $\gamma$-ray emission, quantum corrections due to the speculative memory
burden effect do not modify this mass range. We also investigate the main characteristics
of PBHs which are localized in the bulk.
We show that PBHs localized in the bulk can make all cosmological dark
matter if $10^{11} \alt M_{\rm BH}/{\rm g} \alt 10^{21}$. Finally, we
comment on the black holes that could
be produced if one advocates a space with two boundaries for the dark
dimension.
\end{abstract}

\maketitle

\section{Introduction}

The Swampland  program seeks to distinguish effective theories which
can be completed into quantum gravity in the ultraviolet from those
which cannot~\cite{Vafa:2005ui}. In theory space, the swampland
frontier is drawn
by a family of conjectures classifying the properties that an
effective field theory should have/avoid to enable a consistent completion into quantum gravity. These
conjectures deliver a bridge from quantum gravity to astrophysics,
cosmology, and particle physics~\cite{Palti:2019pca,vanBeest:2021lhn,Agmon:2022thq}.

For example, the distance conjecture (DC) asserts that along infinite distance geodesics there is an infinite tower of states which become exponentially light asymptotically~\cite{Ooguri:2006in}. Connected to the DC is the anti-de
Sitter (AdS) distance conjecture ~\cite{Lust:2019zwm}, which
ties in the dark energy density to the mass scale $m$ characterizing the infinite tower of states,
$m \sim |\Lambda|^\alpha$, in the limit of a small  negative AdS vacuum energy
 with $\alpha$ a positive constant of ${\cal O}
 (1)$. In addition, under the hypothesis that
 this scaling behavior remains valid in dS (or
 quasi dS) space, a large  number of light modes also would pop
 up in the limit $\Lambda \to 0$, with $\Lambda $ being positive.
 
The AdS-DC in de Sitter 
space provides a pathway, called the Dark Dimension scenario~\cite{Montero:2022prj}, to elucidate the origin of the cosmological hierarchy $\Lambda/M_p^{4} \sim 10^{-122}$, because it connects the
size of the compact space $R_\perp$ to the
dark energy scale $\Lambda^{-1/4}$ via
\begin{equation}
  R_\perp \sim \lambda \ \Lambda^{-1/4} \,,
\label{RperpLambda}  
\end{equation}  
where the proportionality factor is estimated to be within the range
$10^{-1} < \lambda < 10^{-4}$. Actually, (\ref{RperpLambda}) derives from
  constraints by theory and experiment. On the one hand, since the associated Kaluza-Klein (KK) 
  tower contains massive spin-2 bosons, the Higuchi
  bound~\cite{Higuchi:1986py} sets an absolute upper limit to the exponent of $\Lambda^\alpha$,
 whereas explicit string calculations
of the vacuum energy~(see
e.g.~\cite{Itoyama:1986ei,Itoyama:1987rc,Antoniadis:1991kh,Bonnefoy:2018tcp})
yield a lower bound on $\alpha$. All in all, the theoretical
constraints lead to $1/4 \leq \alpha \leq 1/2$. On the other hand,
experimental arguments (e.g. constraints on deviations from Newton's
gravitational inverse-square law~\cite{Lee:2020zjt} and neutron star
heating~\cite{Hannestad:2003yd}) lead to the conclusion encapsulated
in (\ref{RperpLambda}): {\it The cosmological hierarchy problem can be
  addressed if there is one extra dimension of
radius $R_\perp$ in the micron range, and the lower bound for $\alpha =
1/4$ is basically saturated~\cite{Montero:2022prj}.} 
The Standard Model (SM) should then be localized on a D-brane transverse to the dark dimension~\cite{Antoniadis:1998ig}.
A theoretical
amendment on the connection between the cosmological and KK mass scales confirms $\alpha =
1/4$~\cite{Anchordoqui:2023laz}. Assembling all this together, we can
further conclude that the KK tower of the new (dark)
dimension opens up at the mass scale $m_{\rm KK} \sim
1/R_\perp$ at the eV range. Within this set-up, the 5-dimensional Planck scale (or species scale
where gravity becomes strong~\cite{Dvali:2007hz,Dvali:2007wp,Cribiori:2022nke,vandeHeisteeg:2023dlw}) is
given by
\begin{equation}
M_* \sim m_{\rm KK}^{1/3}
M_p^{2/3}\, ,
\end{equation}
where $M_p = 1/\sqrt{G_N}$ is the Planck mass, leading to $10^9 \alt
M_*/{\rm GeV} \alt 10^{10}$.

The dark dimension accumulates interesting phenomenology~\cite{Anchordoqui:2022ejw,Anchordoqui:2022txe,Blumenhagen:2022zzw,Anchordoqui:2022tgp,Gonzalo:2022jac,Anchordoqui:2022svl,Anchordoqui:2023oqm,vandeHeisteeg:2023uxj, Noble:2023mfw,Anchordoqui:2023wkm,Anchordoqui:2023tln,Anchordoqui:2023qxv,Law-Smith:2023czn,Anchordoqui:2023etp,Makridou:2023wkb,Obied:2023clp,Anchordoqui:2023woo,Anchordoqui:2024akj,Vafa:2024fpx}. For
example, it was noted by us~\cite{Anchordoqui:2022txe} that 
primordial black holes (PBHs)
with Schwarzschild radius smaller than a micron could be good
dark matter candidates. A generalization to near-extremal black holes
has been carried out in~\cite{Anchordoqui:2024akj}. Complementary to
the PBHs, it was observed in~\cite{Gonzalo:2022jac} that the universal
coupling of the SM 
fields to the massive spin-2 KK excitations of the graviton in the
dark dimension provides an alternative  dark matter contender. The
cosmic evolution of the dark graviton gas is primarily dominated by
``dark-to-dark'' decays, yielding a specific realization of the dynamical dark matter framework~\cite{Dienes:2011ja}. An interesting close
relation between PBHs and the dark gravitons has been pointed out in~\cite{Anchordoqui:2022tgp}.

In this paper we provide new insights on PBHs as dark matter
candidates. The layout is as follows. In Sec.~\ref{sec:2} we provide a
concise overview of black hole evaporation within Hawking's
semiclassical approximation. In Sec.~\ref{sec:3} we explore the impact
of quantum effects, possibly associated to the memory burden, which
could take the evaporation process out of the semiclassical regime by
half-decay
time~\cite{Dvali:2018xpy,Dvali:2020wft,Alexandre:2024nuo}. In
Sec.~\ref{sec:4} we re-assess limits on the abundance of PBHs as dark
matter candidates, focusing on holes localized on the brane during the
evaporation process. In Sec.~\ref{sec:5} we examine the possibility
that bulk PBHs make the dark matter in the universe. In
Sec.~\ref{sec:6} we consider a space
with two boundaries for the dark dimension ~\cite{Schwarz:2024tet}. We study general
phenomenological aspects of this construct, and in particular, we
examine the possibility that the dark matter in the universe is made
of tubular-pancake shape black holes localized on a brane, which is
place at the space boundary.  The
paper wraps up in Sec.~\ref{sec:7} with some conclusions.

\section{Semiclassical Black Hole Evaporation}
\label{sec:2}

In the mid-70's Hawking pointed out that a black hole emits thermal
radiation as if it were a black body, with a temperature inversely
proportional to its mass~\cite{Hawking:1974rv,Hawking:1975vcx}. In
this section we first review the main
properties of Hawking evaporation of Schwarzschild black holes and of
its generalization to dimension $d$. After that we review the
evaporation of near extremal black holes within the semiclassical approximation.

\subsection{Schwarzschild Black Holes}

It is well known that inertial observers in Minkowski space perceive
the vacuum (the absence of particles), whereas observers moving with
uniform proper acceleration $a$ measure a thermal bath (thermal
distribution of particles) of temperature
$T = a/(2\pi)$~\cite{Fulling:1972md,Davies:1974th,Unruh:1976db}. In
special relativity, an observer moving with uniform proper
acceleration through Minkowski spacetime is conveniently described
with Rindler coordinates~\cite{Rindler:1966zz}, which are related to
the standard (Cartesian) Minkowski coordinates by
$x = \rho \cosh \sigma$ and $t = \rho \sinh \sigma$. The line element
in Rindler space is found to be
\begin{equation}
  ds^2 = - \rho^2 d\sigma^2 + d\rho^2 \,,
  \label{rindler}
\end{equation}
where $\rho =1/a$ and where $\sigma$ is related to the observer's
proper time $\tau$ by $\sigma = a \tau$.
Note that the local acceleration diverges as $\rho \to 0$.

By the same token, what inertial Schwarzschild observers measure as vacuum, the
uniformly accelerated ones identify as thermal bath. The group of inertial Schwarzschild observers are the ones falling towards the black
hole, while the uniformly accelerated observers are the ones who
manage to keep constant distance from the event horizon, the
acceleration being there to prevent the gravitational pull. This
implies that black holes should have an approximate Rindler region near the
horizon. More indicatively, consider the Schwarzschild metric 
\begin{equation}
  ds^2 = -  (1-r_s/r) \ dt^2 +  (1-r_s/r)^{-1} \ dr^2 +
  r^2 \ d\Omega_2^2 \,,
\label{Sch-metric}
\end{equation}
where $t$ is the universal time coordinate, $r$ is the circumferential
radius (defined so that the circumference of a sphere of radius $r$ is
$2 \pi r$), $d\Omega_2$ is an interval of spherical solid angle, and
$r_s = 2M_{BH}$ is the horizon radius (in Planck units), with
$M_{\rm BH}$ the black hole mass. The proper distance from the horizon
is given by
\begin{equation}
  \rho  = \int_{r_s}^r \sqrt{g_{rr}(r') } dr' = \int_{r_s}^r \frac{dr'}{\sqrt{1-r_s/r}} 
  =  \sqrt{r(r-r_s)} + r_s\sinh \sqrt{r/r_s -1} \, .
  \label{proper-d}
\end{equation}
Substituting (\ref{proper-d}) into (\ref{Sch-metric}) the Schwarzschild metric can be recast as
\begin{equation}
  ds^2 = - \left(1 - \frac{r_s}{r(\rho)} \right) \ dt^2 + d\rho^2 +
  r^2(\rho) \ d\Omega_2^2 \, .
\end{equation}
For $\delta/r_s \ll 1$, in the near horizon limit, $r=r_s + \delta$, it follows that $\rho = 2
\sqrt{r_s\delta}$. Bearing this in mind, the local metric to lowest
order in $\delta$ is found to be
\begin{equation}
ds^2 = -\frac{\rho^2}{4r_s^2} \ dt^2 + d \rho^2 + r_s^2 \ d\Omega_2^2 \, .
\end{equation}
The $(t,\rho)$ piece of this metric is Rindler space; we can
rescale $t = 2 r_s \sigma$ to make it look exactly like (\ref{rindler}).

In analogy with Rindler space, the near-horizon observer must see the field excited at a local
temperature
\begin{equation}
  T = \frac{a}{2\pi} = \frac{1}{2 \pi \rho} = \frac{1}{4 \pi
    \sqrt{r_s r (1-r_s/r)}} \, .
\end{equation}
The gravitational redshift is given by the square root of the time component of the metric. So for the field theory state to consistently extend, there must be a thermal background everywhere with the local temperature redshift-matched to the near horizon temperature
\begin{equation}
  T(r') = \frac{1}{4 \pi \sqrt{r_s r(1 - r_s/r)}} \
    \sqrt{\frac{1-r_s/r}{1- r_s/r'}} = \frac{1}{4 \pi \sqrt{r_s r
        (1-r_s/r')}} \, .
\end{equation}    
The inverse temperature redshifted to $r'$ at infinity is found to be
\begin{equation}
  \lim_{r' \to \infty} T (r') = \frac{1}{4 \pi \sqrt{r_s r}} \, .
\end{equation}    
For small $\delta$, in the near-horizon limit 
\begin{equation}
  \lim_{r' \to \infty} T (r') = \frac{1}{4 \pi r_s} \, .
\end{equation}
All in all, a field theory defined on Schwarzschild background is in a
thermal state whose Hawking temperature at infinity is given by
\begin{equation}
  T_{\rm H} = \frac{1}{4 \pi \ r_s} \  .
\label{T4d}
\end{equation}  

Armed with the Hawking temperature, we can now calculate the entropy
of the black hole~\cite{Hawking:1976de}. By adding a quantity of heat $dQ$ the change in the
black hole entropy is given by
\begin{equation}
  dS_{\rm BH} = \frac{dQ}{T_{\rm H}} = 8 \pi M_{\rm BH} \ dQ \, =  8
  \pi \ M_{\rm BH} \ dM_{\rm BH} \,,
\end{equation}
where we assumed that the heat energy that enters serves to increase
the total mass. The black hole entropy is then found to be
\begin{equation}
  S_{\rm BH} =  2 \pi \ M_{\rm BH} \ r_s \, .
  \label{S4d}
\end{equation}

Next, in line with our stated plan, we
generalize the previous discussion to dimension $d$. Following~\cite{Giudice:1998ck}, we
conveniently work with the mass scale $M_d = [(2 \pi)^{d-4}/(8
\pi)]^{1/(d-2)}M_*$, where $M_p^2 = M_*^{d-2}  (2 \pi
R_\perp)^{d-4}$~\cite{Antoniadis:1998ig}. Throughout we rely on the probe brane approximation, which ensures that the
only effect of the brane field is to bind the black hole to the
brane. This is an adequate approximation provided $M_{\rm BH}$ is well
above the brane tension, which is presumably of the order of but
smaller than $M_d$. We also assume that the
black hole can be treated as a flat $d$ dimensional object. This
assumption is valid for extra dimensions that are larger than the
$d$-dimensional Schwarzschild radius 
\begin{equation}
  r_s (M_{\rm BH}) \sim \frac{1}{M_d} \left(\frac{M_{\rm BH}}{M_d} \right)^{1/(d-3)}
 \ \left[\frac{2^{d-4} \pi^{(d-7)/2} \Gamma (\frac{d-1}{2})}{d -2}
 \right]^{1/(d-3)} \, ,
 \label{rs}
\end{equation}
where $\Gamma(x)$ is the Gamma
function~\cite{Tangherlini:1963bw, Myers:1986un,Argyres:1998qn}. Using (\ref{rs}) it is easily seen that for dimension $d$, (\ref{T4d}) and (\ref{S4d}) generalize to
\begin{equation}
  T_{\rm H} = \frac{d-3}{4 \pi \ r_s}
\end{equation}
and
\begin{equation}
  S_{\rm BH} = \frac{4 \pi \ M_{\rm BH} \ r_s}{d-2} \,,
\end{equation}
respectively~\cite{Anchordoqui:2001cg}.

In the rest frame of the Schwarzschild black hole, both the average
number~\cite{Hawking:1974rv,Hawking:1975vcx} and the probability
distribution of the number~\cite{Parker:1975jm,Wald:1975kc,Hawking:1976ra} of outgoing particles in each mode
obey a thermal spectrum. However, in the neighborhood of the horizon
the black hole produces an effective potential barrier that
backscatters part of the emitted radiation, modifying the thermal
spectrum. The so-called ``greybody factor'' $\sigma_s$, which controls
the black hole absorption cross section, depends upon the spin of the
emitted particles $s$, their energy $Q$, and $M_{\rm BH}$~\cite{Page:1976df,Page:1976ki,Page:1977um,Kanti:2002nr,Ireland:2023zrd}. The prevailing energies of the emitted particles are
$\sim T_{\rm H} \sim 1/r_s$, resulting in $s$-wave dominance of the
final state.  This implies that the black hole evaporates with equal
probability to a particle on the brane and in the compact
space~\cite{Emparan:2000rs,Dimopoulos:2001hw}. Thereby, the process of evaporation is
driven by the large number of SM brane modes.

At high frequencies ($Qr_s \gg 1$) the greybody factor for each
kind of particle must approach the geometrical optics limit. For
exemplifying simplicity, in what follows we adopt the geometric optics
approximation, and following~\cite{Han:2002yy}, we conveniently write the greybody
factor as a dimensionless constant, $\Gamma_s = \sigma_s/A_4$,
normalized to the horizon surface area
\begin{equation}
  A_4 = 4 \pi \left(\frac{d-1}{2}\right)^{2/(d-3)} \ \frac{d-1}{d-3} \
      r_s^2
    \end{equation}
seen by the SM fields: $\Gamma_{s=0} = 1$, $\Gamma_{s= 1/2} \approx
2/3$, and $\Gamma_{s=1} \approx 1/4$~\cite{Han:2002yy}.

The emission rate per degree of particle freedom $i$ of particles of
spin $s$ with initial total energy between  $(Q, Q+dQ)$  is found to be
\begin{equation}
\frac{d\dot{N}_i}{dQ} = \frac{\sigma_s}{8 \,\pi^2}\,Q^2 \left[
\exp \left( \frac{Q}{T_{\rm H}} \right) - (-1)^{2s} \right]^{-1}
\label{rate}
\end{equation}
The average total emission rate for particle species $i$ is then
\begin{equation}
\dot{N_i} = f \frac{\Gamma_s}{32\,\pi^3} \,
\frac{(d-1)^{(d-1)/(d-3)} \ (d-3)}{2^{2/(d-3)}} \,\Gamma(3) \,
\zeta(3) \,T_{\rm H}\, ,
\label{decayH}
\end{equation}
where $\zeta(x)$ is the Riemann zeta function and
$f=1$ ($f=3/4$) for bosons (fermions)~\cite{Anchordoqui:2002cp}. For
the different spins, the
emission rate can be parametrized by
\begin{equation}
\dot{N}^{s=0}_i \sim 3.7 \times 10^{18}\, \frac{(d-1)^{(d-1)/(d-3)}
\ (d-3)}{2^{2/(d-3)}}\,
\left(\frac{T_{\rm H}}{{\rm MeV}}\right)\,\, {\rm s}^{-1} \,\,,
\end{equation}
\begin{equation}
\dot{N}^{s=1/2}_i \sim 1.8 \times 10^{18}\,
\frac{(d-1)^{(d-1)/(d-3)} \ (d-3)}{2^{2/(d-3)}}\,
\left(\frac{T_{\rm H}}{{\rm MeV}}\right)\,\, {\rm s}^{-1} \,\,,
\end{equation}
and
\begin{equation}
\dot{N}^{s=1}_i \sim 9.2 \times 10^{17}\,\frac{(d-1)^{(d-1)/(d-3)}
  \ (d-3)}{2^{2/(d-3)}}\,
\left(\frac{T_{\rm H}}{{\rm MeV}}\right)\,\, {\rm s}^{-1} \ .
\end{equation}
The rate of change of the black hole mass in the evaporation process
is estimated to be: {\it (i)}~for $d=4$,
\begin{eqnarray}
 \left. \frac{dM_{\rm BH}}{dt}\right|_{\rm evap} & = & -\frac{M_p^2}{30720 \
                                              \pi \ M_{\rm BH}^2} \ \sum_{i}
                                              c_i(T_{\rm H}) \ \tilde f \ \Gamma_s
    \nonumber \\ 
   & \sim & -7.5 \times 10^{-8} \ \left(\frac{M_{\rm BH}}{10^{16}~{\rm
                                                   g}}\right)^{-2} \ \sum_{i}
      c_i (T_{\rm H})  \ \tilde f
 \ \Gamma_s~{\rm g/s}   \,, 
\label{4d}
\end{eqnarray}
and {\it (ii)}~for $d=5$,
\begin{eqnarray}
  \left. \frac{dM_{\rm BH}}{dt}\right|_{\rm evap}
& \sim &  - 9 \ \pi^{5/4} \zeta(4) T_{\rm H}^2 \ \sum_{i} c_i(T_{\rm H}) \ \tilde f
         \ \Gamma_s \nonumber \\
  & \sim &  - 1.3 \times 10^{-12} \ \left(\frac{M_{\rm BH}}{10^{16}~{\rm g}}\right)^{-1} \ \sum_{i}
      c_i (T_{\rm H}) \ \tilde f
\  \Gamma_s~{\rm g/s}  \,,     
\label{5d}
\end{eqnarray}
where $c_i(T_{\rm H})$ counts the number of internal degrees of freedom of particle 
species $i$ of mass $m_i$ satisfying $m_i \ll T_{\rm H}$,  $\tilde f = 1$  $(\tilde f = 7/8)$ for bosons
(fermions), and where for $d=5$, we have taken $M_* \sim 10^{10}~{\rm GeV}$ as expected in the
dark dimension scenario~\cite{Montero:2022prj}. A direct comparison of (\ref{4d}) and
(\ref{5d}) shows that for equal masses, five-dimensional (5D) Schwarzschild black holes evaporate
slower than their 4D cousins. This is because 5D black holes are
bigger and colder. Indeed, using (\ref{4d}) a straightforward
calculation shows that 4D black holes with $M_{\rm BH} \agt 5 \times
10^{14}~{\rm g}$ can survive the age of the universe, whereas for the
dark dimension scenario we obtain
\begin{equation}
  \tau_H \sim 13.8 \ \Bigg(\frac{M_{\rm BH}}{10^{12}~{\rm g}} \Bigg)^2 \ \left(\frac{6}{\sum_{i}
      c_i (T_s) \ \tilde f
      \  \Gamma_s}\right)~{\rm Gyr} \, .
\label{lifetime}  
\end{equation}
A discussion of a 4D black hole evaporation entering the 5D regime
together with details of the 4D to 5D transition is provided in the Appendix.

\subsection{Near-Extremal Black Holes}

Next, we consider the Reissner–Nordstr\"om black hole in
Einstein-Maxwell gravity, which is specified by its ADM mass $M_{\rm BH}$ and
electric charge $Q_{\rm BH}$. The Einstein-Maxwell action in dimension $d$ is given by,
\begin{equation}\label{actem}
S_{\rm EM}=-\frac{1}{16 \pi G}\int_M d^dx\ \sqrt{-g} \ \Bigl(R-F^2 \Bigr) \
,
\end{equation} 
where $F=dA$ and
\begin{equation}
A=-\sqrt{\frac{d-2}{2(d-3)}} \ \frac{q}{r^{d-3}} \ dt \ .
\end{equation}
The $d$-dimensional Reissner-Nordstr\"om metric can be written as
\begin{equation}
ds^2 = -u(r) \ dt^2 + u^{-1} (r) \ dr^2 + r^{2} \ d\Omega_{d-2}^2\,,
\end{equation}
where
\begin{equation}
u(r) = 1 - \frac{m}{r^{d-3}} + \frac{q^2}{r^{2(d-3)}} 
\end{equation}
is the blackening function and
\begin{equation}
  d\Omega^2_{d-2} = d \chi_2^2 + \prod_{i=2}^{d-2} \sin^2 \chi_i \
  d\chi^2_{i+1}
\end{equation}  
is the metric of a $(d-2)$-dimensional unit sphere, and where $m$ is
related to the ADM mass $M_{\rm BH}$ and $q$ to the charge $Q_{\rm
  BH}$ of the black hole by
\begin{equation}
  M_{\rm BH} = \frac{d-2}{16 \pi G} \ \omega_{d-2} \ m
\end{equation}  
and
\begin{equation}
Q_{\rm BH} =\frac{\sqrt{2(d-2)(d-3)}}{8\pi G}\,\omega_{d-2}\ q \ ,
\end{equation} 
with
\begin{equation}
\omega_{d-2}= \frac{2\pi^{(d-1)/2}}{\Gamma\left(\frac{d-1}{2}\right)} 
\end{equation}
the volume of the unit $(d-2)$-sphere~\cite{Chamblin:1999tk}. The
black hole has two horizons located at the zeroes of the blackening function,
$u(r_\pm) = 0$, yielding
\begin{equation}
r_\pm =  \left[(M_{\rm BH} + c)/M_d^{d-2}\right]^{1/(d-3)} \,,
\end{equation}
where following~\cite{Cribiori:2022cho} we conveniently introduce the extremality parameter
\begin{equation}
  c = \sqrt{M_{\rm
      BH}^2 - Q_{\rm BH}^2 M_d^{d-2}} \sim M_{\rm BH}
  \sqrt{\beta/S_{\rm BH}} \,,
\label{c}
\end{equation}  
and where we have
been cavalier on unimportant numerical factors taking  $m \sim 2M_{\rm
  BH}/M_d^{d-2}$ and $q \sim Q_{\rm BH}/M_d^{d-2}$. To avoid naked
singularities, we must require $M_{\rm BH}^2  \geq Q_{\rm BH}^2 M_d^{d-2}$.

Now, we define the temperature,
\begin{equation}
  T_{\rm ne} \sim \frac{\beta^{1/2} T_{\rm H}}{S_{\rm BH}^{1/2}} \,,
\label{Tne}
\end{equation}
where $\beta$ is a factor of order-one that controls the differences
between $M_{\rm BH}$ and $Q_{\rm BH}$~\cite{Basile:2024dqq}. By
inspection of (\ref{c}) and (\ref{Tne}) we can see that near-extremal black holes are extremely cold due to the fact
that $c/M_{\rm BH}$, which quantifies the near-extremality, is extremely small because of the large entropy~\cite{Anchordoqui:2024akj}.\\

\section{Effects of Memory Burden on Black Hole Evaporation}

\label{sec:3}

It was recently suggested ~\cite{Dvali:2018xpy,Dvali:2020wft,Alexandre:2024nuo} that quantum effects (e.g. memory burden)
would take the evaporation process out of the semiclassical regime
after about half-decay time has elapsed. It is this that we now turn to study.
Actually,
the black hole evaporation estimates presented in the previous
section, based on the
semiclassical approximation, rely on the assumption of
self-similarity. Namely, we have assumed that in the course of
evaporation, a black hole gradually shrinks in size while maintaining
the standard semi-classical relations between its parameters, such as
its mass, the Schwarzschild radius, and the Hawking
temperature. However, there have been some objections to this
viewpoint, and actually arguments have been put forward suggesting the self-similarity assumption
is  inconsistent over the long time-scales of the evolution.

The breakdown of self-similarity is deep-rooted on the nature of
entanglement. For a cavity emitting black body radiation, early on photon
emission is entangled with atomic states in the
cavity. However, once half the energy in the cavity is emitted
subsequent radiation emitted is entangled with radiation emitted
earlier. As a result the entanglement entropy increases to some
maximum, at about half the energy emitted, and then declines. A black
hole may experience a similar behavior, because Hawking radiation is emitted from an
entangled pair of photons or electron positron pairs. The Page time is
defined by the condition that the mass of a black hole has decreased
to half its original value via Hawking radiation, $\tau_{\rm half}
\sim M_{\rm BH}/2$~\cite{Page:1993wv,Page:2013dx}. Within this time $r_s \to r_s/2$ and $S_{\rm BH}
\to S_{\rm BH}/4$. It is at time $\sim \tau_{\rm half}$ when an
observer that has configured a black hole with a set of known
 states on the horizon might find that they have been randomized beyond what can
 be recovered. In other words,  information may remain encoded inside the
 black hole because the emitted radiation is thermal in
 character. However, after $\tau_{\rm half}$, the remaining black hole
 has only $1/4$ of its initial entropy and so much less information
 storage capacity. A far reaching proposal that can accommodate these ideas and describe the
black hole evaporation for times $ > \tau_{\rm half}$ is
the so called  {\it memory burden} effect, which insinuates that Hawking evaporation is slowed
down by further $n$ powers of the entropy $S_{\rm BH}$, after the Page
time has elapsed~\cite{Dvali:2018xpy,Dvali:2020wft,Dvali:2021byy,Alexandre:2024nuo}.

It is not clear whether the memory burden effect still needs further
justification, and whether the black hole lifetime already gets
increased after Page time, but herein we will explore its spectacular 
phenomenological consequences. Before proceeding, we pause to note that
memory burden  holds in any dimension and herein we assume 
that the behavior of $t_{\rm half}$ is roughly universal for any dimension.

Bearing this in mind, a comparison of the scaling behavior of the different rates of
particle emission is of interest. The scaling behavior of the semiclassical
decay rate ($\dot N_i \equiv \Gamma_{\rm H}^{\{0\}}, \ \forall i$) of
a Schwarzschild black hole in dimension $d$ follows from  (\ref{decayH}). Neglecting the
effect of grebody factors and mass thresholds we have
\begin{equation}
   \Gamma_{\rm H}^{\{0\}} \sim T_{\rm H} \sim
 M_d \  S_{\rm BH}^{-1/(d-2)} \sim M_d \ \left(\frac{M_{\rm
       BH}}{M_d}\right)^{-1/(d-3)} \, .
\end{equation} 
The quantum decay rate has an
additional suppression (compared to the Hawking decay rate), 
\begin{equation}
   \Gamma_{\rm H}^{\{n\}} \sim \frac{T_{\rm H}}{S_{\rm BH}^n} \sim M_d \ S_{\rm
     BH}^{(-1-nd+2n)/(d-2)} \sim M_d \ \left(\frac{M_{\rm BH}}{M_d}\right)^{(-1-nd+2n)/(d-3)} \,,
 \end{equation}
and the suppression
factor, $1/S_{\rm BH}^n \sim (M_{\rm BH}/M_d)^{(-nd+2n)/(d-3)}$, is
dimension-dependent. Throughout, $n$ is a non-negative integer parametrizing the quantum suppression when the
black hole enters the memory burden phase.

\begin{figure}[htb!]
  \begin{minipage}[t]{0.48\textwidth}
    \postscript{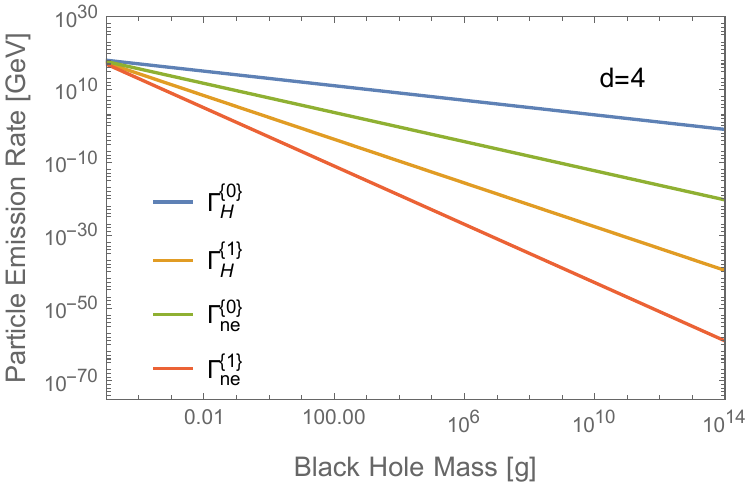}{0.9}
  \end{minipage}
\begin{minipage}[t]{0.48\textwidth}
    \postscript{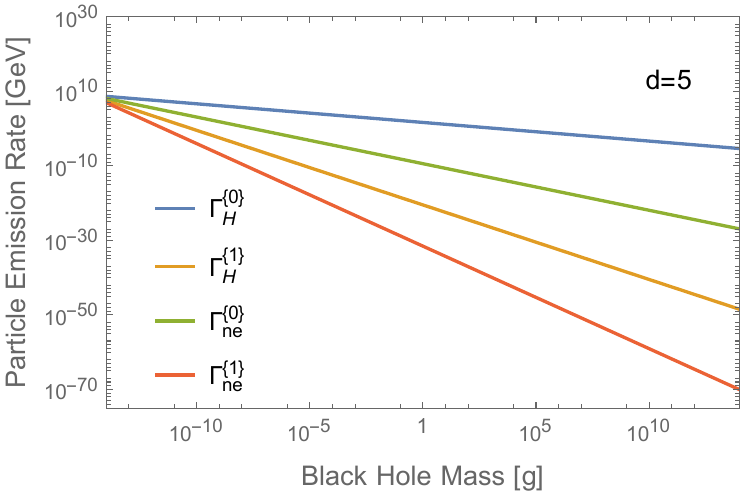}{0.9}
  \end{minipage}
  \caption{Particle emission rate of
    Schwarzschild, near extremal, and quantum
    black holes; for $d=4$ (left) and $d=5$ (right). \label{fig:1}}
  \end{figure}

For near-extremal black holes, the semiclassical decay rate in
dimension $d$ is suppressed  compared to the Schwarzschild decay rate
by a factor $1/S_{BH}^{1/2}$~\cite{Anchordoqui:2024akj} 
\begin{equation}
   \Gamma_{\rm ne}^{\{0\}} \sim \frac{\beta^{1/2} \ T_{\rm H}}{S_{\rm BH}^{1/2}}
   \sim \beta^{1/2} \ M_d \ S_{\rm BH}^{-d/(2d-4)} \sim \beta^{1/2} \ M_d
   \ \left(\frac{M_{BH}}{M_d}\right)^{-d/(2d-6)} \, .
\end{equation}
The quantum decay
rate of a near-extremal black hole in
    dimension $d$ gets again an additional suppression compared to the near-extremal Hawking decay rate by a factor $1/S_{BH}^n$,
\begin{equation}
   \Gamma_{\rm ne}^{\{n\}}  \sim  \frac{\beta^{1/2}  T_{\rm H}}{S_{\rm
       BH}^{n+1/2}} \sim
                              \beta^{1/2}  M_d  S_{\rm BH}^{(-d/2-nd+2n)/(d-2)} 
  \sim  \sqrt{\beta} M_d
  \left(\frac{M_{\rm BH}}{M_d}\right)^{(-d/2-nd+2n)/(d-3)} \, .
\end{equation}
In Fig.~\ref{fig:1} we show a comparison of the various decay
rates. For large masses (or equivalently large entropies), the quantum
decay rates are suppressed by several orders of magnitude. The
difference, of course, is reduced with decreasing entropy, and all the decay rates
become equal at the minimum black hole mass $M_{\rm BH, min} \sim
M_d$.

In summary, Hawking's semiclassical approximation applied to 5D black holes
perceiving the dark dimension scenario gives very robust predictions and lead to reliable
conclusions. More spectacular results can be obtained adopting the
black hole portrait endowed with the memory burden. However, the
theoretical underpinning and the precise implementation of the memory burden is
rather speculative. In particular via memory burden the black hole decay rate  gets dramatically changed already after Page time, whereas it also could be conceivable that quantum effects affect the black hole decay rate at much later
time scales. Therefore, it will be very interesting to provide
further theoretical evidence for memory burden associated closer to
the entanglement island
approach~\cite{Ryu:2006bv,Hubeny:2007xt,Lewkowycz:2013nqa,Penington:2019npb,Almheiri:2019psf,Almheiri:2019hni,Penington:2019kki,Almheiri:2019qdq,Hashimoto:2020cas,Chen:2020jvn,Bousso:2023kdj}.

\section{Primordial Black holes as dark matter}
\label{sec:4}

Since the late 60's there have been speculation suggesting that black
holes could be formed from the
collapse of large amplitude fluctuations in the early
universe~\cite{Zeldovich:1967lct,Hawking:1971ei,Carr:1974nx,Carr:1975qj}. An
order of magnitude estimate of $M_{\rm BH}$ can be obtained by
equating the scaling of the cosmological energy density with time $t$
in the radiation dominated epoch,
\begin{equation}
  \rho \sim M_p^2/t^2 \,,
\label{rhouno}
\end{equation}
to the required density in a region of mass
$M_{\rm BH}$ which is able to collapse within its Schwarzschild radius
\begin{equation}
\rho \sim  M_p^6/M_{\rm BH}^2 \, .
\label{rhodos}
\end{equation}
Linkening (\ref{rhouno}) and (\ref{rhodos}) gives the idea that at production PBHs would
have roughly the cosmological horizon mass~\cite{Carr:2020xqk}
\begin{equation}
M_{\rm BH} \sim  t M_p^2 \sim 10^{15}
\left(\frac{t}{10^{-23}~{\rm s}}\right)~{\rm g} \, .
\label{yieldrho}
\end{equation}
Now, it is straightforward to see that a black hole would have the Planck mass ($M_p
\sim 10^{-5}~{\rm g}$) if it formed at the Planck time
($10^{-43}~{\rm s}$), $1~M_\odot$ if it formed at the QCD epoch
($10^{-5}~{\rm s}$), and $10^{5} M_\odot$ if it formed at $t \sim
1~{\rm s}$, comparable to the mass of the holes thought to reside in
galactic nuclei. The back-of-the-envelope calculation yielding (\ref{yieldrho}) hints that PBHs could span an enormous mass
range. Even though the spectrum of masses of these PBHs is yet to see
the light of day, on cosmological scales they would behave like a typical cold dark
matter particle.

First of all, an all-dark-matter interpretation in terms of
PBHs, namely  the possible mass range for PBHs is constrained by the requirement that they must have survived up today, i.e. PBHs must have lived longer than the age of the universe.
Additional severe constraints are provided by several
observations~\cite{Carr:2020xqk,Green:2020jor,
  Villanueva-Domingo:2021spv,LISACosmologyWorkingGroup:2023njw}. 
  To be specific, the extragalactic
$\gamma$-ray background~\cite{Carr:2009jm}, the cosmic microwave
background (CMB)~\cite{Clark:2016nst}, the 511~keV $\gamma$-ray
line~\cite{DeRocco:2019fjq,Laha:2019ssq,Dasgupta:2019cae,Keith:2021guq},
EDGES 21-cm signal~\cite{Mittal:2021egv}, and the MeV Galactic diffuse emission~\cite{Laha:2020ivk,Berteaud:2022tws,Korwar:2023kpy} constrain evaporation of black holes with masses $\lesssim 10^{17}~{\rm g}$,
whereas the non-observation of microlensing events by MACHO~\cite{Macho:2000nvd}, EROS~\cite{EROS-2:2006ryy},
  Kepler~\cite{Griest:2013aaa}, Icarus~\cite{Oguri:2017ock},
  OGLE~\cite{Niikura:2019kqi} and Subaru-HSC~\cite{Croon:2020ouk} set
  an upper limit on the black hole abundance for masses $M_{\rm BH} \gtrsim
  10^{21}~{\rm g}$.

Before proceeding, we pause and call attention to a captivating
coincidence:
\begin{equation}
{\tt size \ of \ the  \ dark \ dimension
}  \sim {\tt wavelength \ of \ visible \ light} \,,
\label{stun}
\end{equation}
which implies that the
  Schwarzschild radius of 5D black holes is well below the wavelength
  of light. For point-like lenses, this is precisely the critical length where
  geometric optics breaks down and the effects of wave optics suppress
  the magnification, obstructing the sensitivity to 5D PBH
  microlensing signals~\cite{Croon:2020ouk}.
  So 5D PBHs escape these microlensing constraints; at the same time, as pointed out in~\cite{Anchordoqui:2022txe},
  they are longer lived than their 4D counter parts, and this makes them more promising all-dark-matter candidates.

In light of the stunning coincidence (\ref{stun}), hereafter we will now
  focus on the bounds imposed by the extragalactic
$\gamma$-ray background~\cite{Carr:2009jm}, the CMB
spectrum~\cite{Clark:2016nst}, the 511~keV $\gamma$-ray
line~\cite{DeRocco:2019fjq,Laha:2019ssq}, and the MeV Galactic diffuse emission~\cite{Laha:2020ivk,Berteaud:2022tws,Korwar:2023kpy} which directly constrain the black hole decay
rate. As previously noted, these constraints place an upper limit on the
dark matter fraction $f_{\rm PBH}$ composed of PBHs.

Before proceeding, we pause to note that femtolensing of
  cosmological gamma-ray bursts (GRBs) could be used to search for PBHs in the mass range
  $10^{17} \alt M_{\rm BH}/{\rm g} \alt
  10^{20}$~\cite{Gould:1992}. The lack of
  femtolensing detection by the {\it Fermi} Gamma-ray Burst Monitor was used
  to constrain PBHs with mass in the range $10^{17.7} < M_{\rm
    BH}/{\rm g} \alt 10^{20}$ to
  contribute no more than 10\% to the total dark matter
  abundance~\cite{Barnacka:2012bm}. However, the validity of this GRB constraint has been called
  into question,
  because it is based on the assumption that the gamma-ray source is
  point-like in the lens plane~\cite{Katz:2018zrn}. The non-pointlike nature of GRBs imply that most of them are too big when projected onto the
  lens plane to yield meaningful femtolensing limits. Indeed, only a
  small GRB population with very fast variability might be suitable for PBH
  searches. When this systemtic effect is taken into consideration to
  decontaminate the {\it Fermi} 
  sample, the GRB constraint on PBHs becomes obsolete. A sample of 100
  GRBs with transverse size $10^9~{\rm cm}$ would be needed to probe
  the 5D PBHs~\cite{Katz:2018zrn}.

Black hole evaporation has two main channels
contributing to the (diffuse) isotropic $\gamma$-ray flux: {\it (i)} direct photon
emission and {\it (ii)} positron emission (these positrons annihilate with
thermal electrons producing an $X$-ray background at energies around
and below 511~keV). The diffuse photon emission consists of a
contribution from PBH from extragalactic structures at all redshifts. Limits on $f_{\rm PBH}$ are obtained by  demanding that the emission from
PBH evaporation does not exceed the $X$-ray and soft
$\gamma$-ray isotropic fluxes, neither the Galactic intensity
associated with the line-of-sight integrated Navarro-Frenk-White dark matter density profile~\cite{Navarro:1995iw}.
Observations of the diffuse $\gamma$-ray emission in the direction of
the Galactic center offer the best and most solid constraints of
$f_{\rm PBH}$~\cite{Korwar:2023kpy}.

\begin{figure}
\postscript{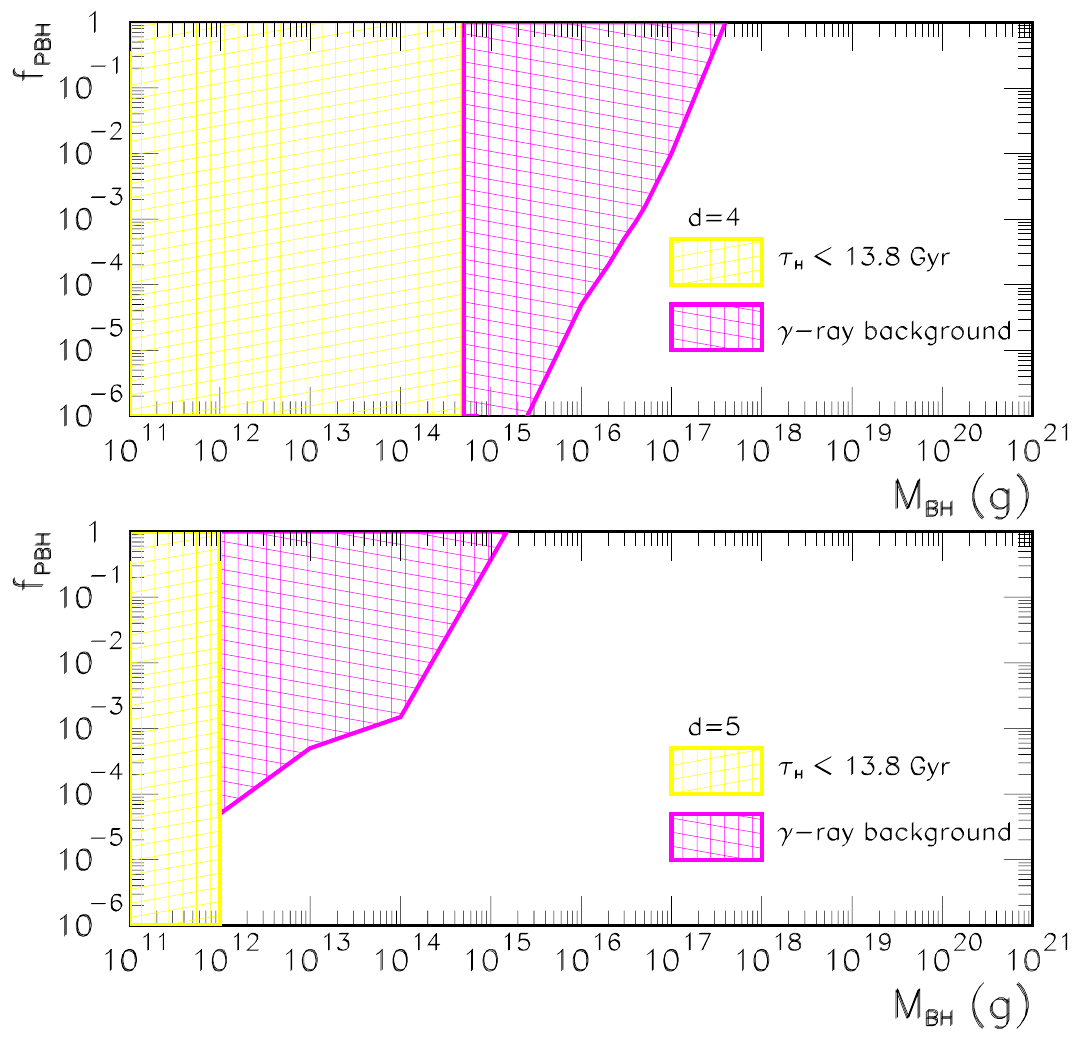}{0.9}
\caption{Contraints on $f_{\rm PBH}$ as a function of the PBH mass $M_{\rm BH}$, assuming a monochromatic mass
  function. The yellow area indicates the region where black holes do
  not survive the age of the universe. The region shaded in magenta
  corresponds to the upper limits from the extragalactic $\gamma$-ray
  background~\cite{Carr:2009jm}, the CMB
  spectrum~\cite{Clark:2016nst}, the 511~keV $\gamma$-ray
line~\cite{DeRocco:2019fjq,Laha:2019ssq}, and the MeV Galactic diffuse emission~\cite{Laha:2020ivk,Berteaud:2022tws,Korwar:2023kpy}. The upper panel shows the results for $d=4$, and the
  lower panel for $d=5$.}
\label{fig:2}
\end{figure}

The upper panel of
Fig.~\ref{fig:2} shows the constraints on the PBH fraction in the mass
range of interest for $d=4$. The concomitant constraints for $d=5$, can
be estimated by a simple rescaling procedure. The central idea to
decipher the change of scale is that for a given photon energy, or equivalently a given
Hawking temperature, it is reasonable to expect a comparable limit on
$f_{\rm PBH}$ for both $d=4$ and $d=5$, i.e.,
\begin{equation}
  \rho_{\rm PBH}^{4D} (T_{\rm H}) \sim \rho_{\rm PBH}^{5D}(T_{\rm H}) \, .
\end{equation}  
Note that at given $T_{\rm H}$, the required PBH number
density in 5D is larger than the one in 4D
\begin{equation}
  n_{\rm PBH}^{4D} (T_{\rm H}) < n_{\rm PBH}^{5D} (T_{\rm H}) \,,
\end{equation}
but this is
automatically  compensated by the fact that at fixed Hawking temperature the 5D
black holes have smaller $M_{\rm BH}$ than those in 4D, i.e.,
\begin{equation}
  M_{\rm BH}^{4D} (T_{\rm H}) > M_{\rm BH}^{5D} (T_{\rm H}) \, ,
\end{equation}
yielding equal density.

For example, in $d=4$, the Hawking temperature is related to the mass of the black hole by~\cite{Keith:2021guq}
\begin{equation} 
T_H = \frac{M_p^2}{8 \pi M_{\rm BH}} \sim 
\, \bigg(\frac{M_{\rm BH}}{10^{16} \, {\rm g}} \bigg)^{-1}~{\rm MeV}\,,
\end{equation}
whereas in $d=5$ the Hawking temperature mass relation
is found to be
\begin{equation}
  T_H  \sim 1/r_s \sim 
    \left(\frac{M_{\rm BH}}{10^{12}~{\rm g}}\right)^{-1/2}~{\rm MeV} \, ,
\label{tempes}
\end{equation}
where we have taken $M_d \sim 10^{10}~{\rm GeV}$.\footnote{We adopt the
highest value of $M_d$ to remain conservative in the estimated bound on $f_{\rm PBH}$.} This means that since
$f_{\rm PBH} \alt 5 \times 10^{-5}$ in $d=4$ for $M_{\rm BH} \sim
10^{16}~{\rm g}$, our reasoning indicates that $f_{\rm PBH}
\alt 5 \times 10^{-5}$ for $M_{\rm BH} \sim
10^{12}~{\rm g}$ in $d=5$. The complete rescaling of $f_{\rm PBH}$ for the
$M_{\rm BH}$ range of interest is displayed in the lower panel of Fig.~\ref{fig:2}.

In Sec.~\ref{sec:3} we have seen that near-extremal black holes are extremely cold. For example, a black hole of $M \sim 10^5~{\rm g}$ has a temperature $T_H \sim 4~{\rm GeV}$. Substituting these figures into
      (\ref{4d}) we find that the Hawking lifetime of a $10^5~{\rm g}$
      Schwarzschild black hole is $\tau_H \sim 4
      \times 10^{-5}~{\rm yr}$. For a near extremal black hole of the
      same mass, using (\ref{Tne}) we find that its temperature would
      be $T_{\rm ne} \sim 10^{-5} \sqrt{\beta}~{\rm
        eV}$, and so using  (\ref{lifetime}) its lifetime $\tau_{\rm ne} \sim 15/\sqrt{\beta}~{\rm
        Gyr}$. Note that the energy of the emitted particles by the
      near-extremal black hole is well
      below the peak of the CMB photons. We can conclude that if there were 5D primordial near-extremal black holes in nature, then a PBH all-dark-matter
interpretation would be possible in the mass range 
      \begin{equation}
        10^5 \sqrt{\beta} \alt M/{\rm g} \alt 10^{21} \, .
\end{equation}
Of course, by tuning the $\beta$ parameter we can have a PBH
all-dark-matter interpretation with very light 5D black
holes.\footnote{The possibility of 4D near-extremal black holes making
  the dark matter was suggested in~\cite{deFreitasPacheco:2023hpb}.}

\begin{figure}
\postscript{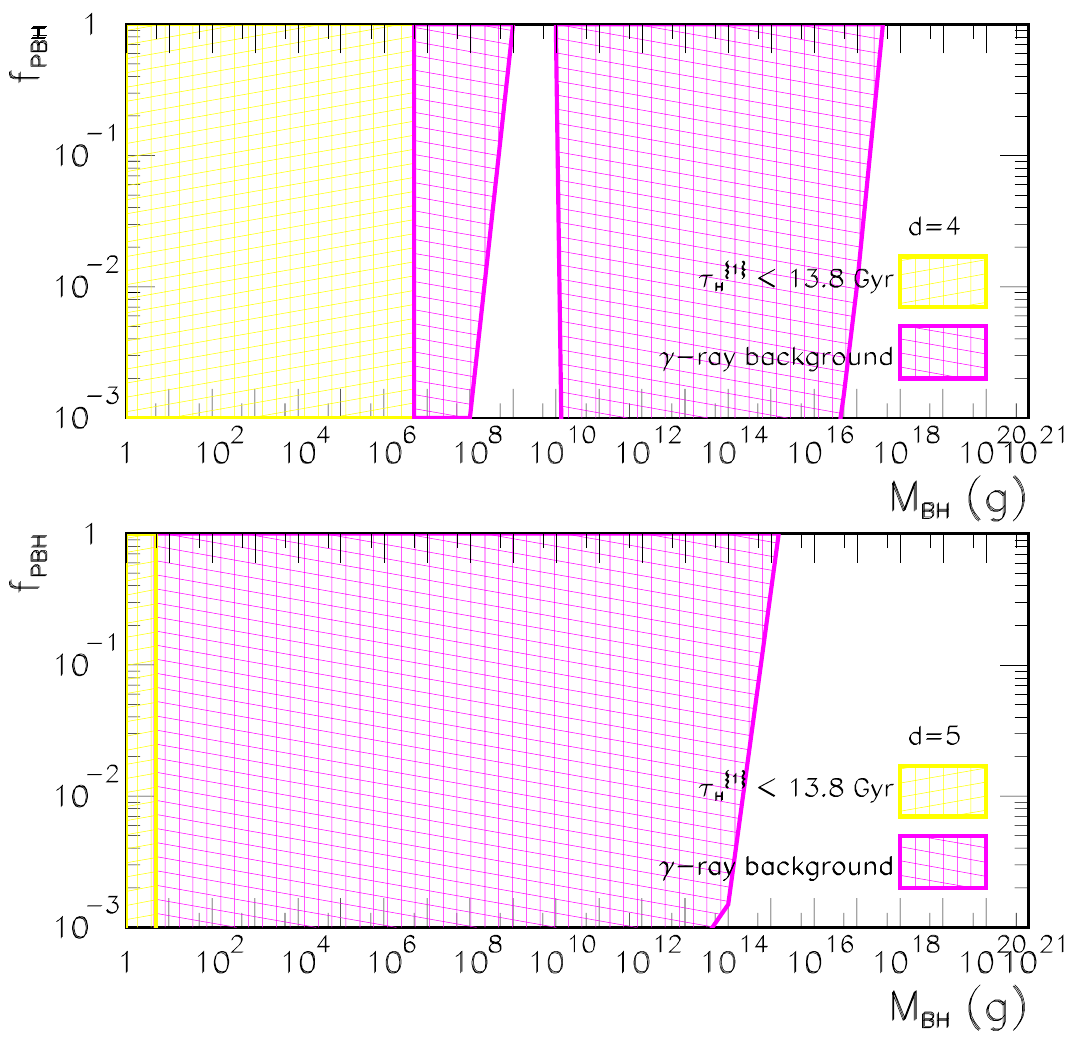}{0.9}
\caption{Contraints on $f_{\rm PBH}$ as a function of the PBH mass $M_{\rm BH}$, assuming a monochromatic mass
  function. The yellow area indicates the region where black holes do
  not survive the age of the universe, i.e. $\tau_{\rm H}^{\{1\}} <
  13.8~{\rm Gyr}$. The region shaded in magenta
  corresponds to the upper limits from the extragalactic $\gamma$-ray
  background~\cite{Carr:2009jm}, the CMB
  spectrum~\cite{Clark:2016nst}, the 511~keV $\gamma$-ray
line~\cite{DeRocco:2019fjq,Laha:2019ssq}, and the MeV Galactic diffuse emission~\cite{Laha:2020ivk,Berteaud:2022tws,Korwar:2023kpy}. The upper panel shows the results for $d=4$, and the
  lower panel for $d=5$.}
\label{fig:3}
\end{figure}

Next, we explore how the memory burden effect impacts
the mass range of a PBH all-dark-matter interpretation.  Memory burden dramatically enhances the black hole life time and implies that a-priori the bounds from the age of the universe are much milder
than compared to those from the semiclassical black hole decay. However, photon emission will put additional constraints, which we will discuss in the following.
To be specific,
we assume only
two phases in the evaporation process: semiclassical evaporation up to
$\tau_{\rm half}$ followed by a quantum regime characterized by $n=1$. We can duplicate the
procedure adopted above to derive the allowed mass range for a PBH
all-dark-matter interpretation. For $d=4$, the memory burden effect
opens a new window in the mass range $10^9 \alt M_{\rm BH}/{\rm g}
\alt 10^{10}$~\cite{Thoss:2024hsr}. This corresponds to PBHs with
temperatures $1 \alt T_H/{\rm TeV} \alt 10$. The corresponding 5D mass
range for the same 
temperature interval is $10^{-2} \alt M_{\rm BH}/{\rm g} \alt 1$. Imposing the
$S_{\rm BH}$ enhancement on (\ref{lifetime}), we find that 
\begin{equation}
  \tau_{\rm H}^{\{1\}} \sim 4 \times 10^{17} \left(\frac{M_{\rm BH}}{5~{\rm
      g}} \right)^2~{\rm s} \, ,
\end{equation}
i.e., black holes could survive the age of the universe if $M_{\rm BH}
\agt 5~{\rm g}$. Bounds on the black hole abundance as a function of
the initial black hole mass are encapsulated in Fig.~\ref{fig:3}. We
conclude that within the dark dimension scenario the memory burden effect with $n=1$
does not open any new window that could allow an all-dark-matter
interpretation composed of primordial black holes.

\section{Black Holes Localized in the Bulk}
\label{sec:5}

Hitherto, we have assumed that 5D black holes stay attached to the
brane during the evaporation process. In this section we relax this
assumption and allow them wander off into the bulk. Without knowing
more details of the bulk and brane theory it is not worth considering to calculate the probability of
such wandering in detail. However, we can assume that the holes are
out of the brane-world and study the evaporation effects of these bulk
PBHs. Furthermore, it is always possible that the PBHs are
produced in the bulk to start with. This situation will be more
appealing within the proposal introduced elsewhere~\cite{Anchordoqui:2022svl,Anchordoqui:2023etp}, in which we
postulated that the dark dimension
may have undergone a uniform rapid expansion, together with the
three-dimensional non-compact space, by regular exponential inflation
driven by an (approximate) higher dimensional cosmological constant. If this were the case, then primordial fluctuations during inflation of the compact space could lead to the
production of black holes in the bulk. In this
section we then assume
that PBHs are
localized or propagating in the bulk.
 
Bulk black holes live longer than those attached to the brane. This
is because KK modes are excitations in the full transverse space and so their
overlap with small (higher dimensional) black holes is suppressed by the
geometric factor $(r_s/R_\perp)^{d-4}$ relative to the brane fields. This
geometric suppression precisely compensates for the enormous number of
modes and the total KK contribution is only of same order as that from
a single brane field~\cite{Emparan:2000rs}. Actually, greybody factors suppress graviton
emission when compared to fermions and gauge bosons, and hence bulk
black holes which do not have access to the brane degrees of freedom
are expected to live longer. In
addition, since there is no emission on the brane the bounds due to
photon evaporation can be avoided. This implies that PBHs localized in
the bulk can provide an all-dark-matter interpretation if
\begin{equation}
  10^{11} \alt M_{\rm
    BH} /{\rm g} < 10^{21} \,,
\label{cuarentaycuatro}
\end{equation}
where we have remained conservative, and following~\cite{Han:2002yy} we assumed that the ratio of the emitted flux into a single
brane field over a single bulk field is
 roughly a factor of two.

We note that within the black hole memory burden an all dark-matter
interpretation is almost unconstrained, and so the allowed mass range estimate
given 
in (\ref{cuarentaycuatro}) can be extended down to  $1 \alt
M_{\rm BH}/{\rm g} \alt 10^{21}$. Actually, the minimum black hole
mass can be further reduced by advocating a
multiple-phase black hole evolution, such that in each phase there is
a $1/S_{\rm BH}$ increase in the memory burden suppression factor of the decay rate.

\section{The dark dimension as a space with two boundaries}
\label{sec:6}

Very recently, it was conjectured that
 the dark dimension can also be viewed as a line interval with end-of-the-world
9-branes attached at each
end~\cite{Schwarz:2024tet}. In this section, we briefly comment on the
impact that this thought-provoking viewpoint could have on
phenomenological aspects of the
dark dimension scenario.

Actually, the line interval along the dark dimension $y$ can also be  understood as a semi-circular dimension
endowed with $S^1/\mathbb{Z}_2$
symmetry. This symmetry has radical consequences for models in which neutrino masses originate in 
three 5D Dirac fermions  $\Psi_\alpha$, which are singlets under the SM gauge symmetries and interact in our brane with the three active left-handed neutrinos
  $\nu_{\alpha L}$ in a way that conserves lepton number, where the indices $\alpha = e, \mu,\tau$ denote the
  generation~\cite{Dienes:1998sb,Arkani-Hamed:1998wuz,Dvali:1999cn}. In
  the Weyl basis each Dirac field can be decomposed into two two-component
spinors $\Psi_\alpha \equiv
(\psi_{\alpha L},\psi_{\alpha R})^T$.
Now, the $\mathbb{Z}_2$ symmetry of $S^1/\mathbb{Z}_2$ contains $y$ to $-y$ which acts as chirality ($\gamma_5$) on spinors.
This implies that one of the two-component Weyl
spinors, say $\psi_{\alpha R}$, would be even under $\mathbb{Z}_2$,
while the other spinor $\psi_{\alpha L}$ would be odd. If $\nu_{\alpha
  L}$ are 
restricted to the brane located at the fixed point $y = 0$, then
$\psi_{\alpha L}$ vanishes at this point and so the  coupling is
between $\nu_{\alpha L}$ and $\psi_{\alpha R}$. As we have shown
elsewhere~\cite{Anchordoqui:2023woo},  the model remains
consistent with experiment because only the $\psi_{\alpha R}$ degrees of freedom contribute to the effective number of
equivalent neutrino species
$N_{\rm eff}$.

Two different types of PBHs can be produced if the dark dimension is
seen as space with two
boundaries: {\it
  (i)}~5D Schwarzschild black holes that propagate freely (or else are
localized) in the bulk
and {\it (ii)}~tubular-pancake shape black holes, which are bound to
a brane localized at the boundary of space. The second type of black holes would resemble the black cigars described
in~\cite{Chamblin:1999by}, and therefore far away from the space
boundary the black hole metric on the end-of-the-world brane would be approximately Schwarzschild. In the spirit of~\cite{Anchordoqui:2002fc}, we argue that the
phenomenology of these black cigars is similar to that of the 5D Schwarzschild
black holes discussed throughout this paper. This implies that
primordial black
cigars could provide an all-dark-matter interpretation for masses in
the range
$10^{15} < M_{\rm BH} /{\rm g} < 10^{21}$; see Fig.~\ref{fig:2} for
details. We have estimated in the previous section that to accommodate the
observed dark matter density Schwarzschild 
black holes living in the bulk should have masses in the range $10^{11} \alt M_{\rm
    BH} /{\rm g} < 10^{21}$.
 
We have stressed in the Introduction that the dark graviton gas proposal requires KK
excitations which are unstable~\cite{Gonzalo:2022jac}. Actually, there are strong bounds on the changing dark matter
density which imply that the decays to lighter gravitons of the KK tower must not lose much mass to kinetic energy~\cite{Obied:2023clp}. A typical violation of KK
quantum number $\delta_n$ cannot be too large and requires $\delta_n \sim {\cal
  O}(1)$. The proposal entertained in~\cite{Schwarz:2024tet} provides a profitable arena to accommodate the
dark graviton gas cascade, because
$S^1/\mathbb{Z}_2$ automatically breaks the $U(1)$ isometry associated to the KK
 momentum conservation. Indeed, for $S^1/\mathbb{Z}_2$ there are two level sources of KK momentum violation:
\begin{itemize}
\item KK momenta are conserved modulo 2;
\item KK states decay to brane states with a coupling that falls off
  exponentially at large KK number $n>M/m_{\rm KK}$ with $M$ the brane
  width. This of course feeds back to a momentum violation among KK
  modes by a loop of brane states. 
\end{itemize}
Now, the KK momentum violation can leave a discrete $\mathbb{Z}_2$
symmetry acting as the KK number parity. This symmetry can be, in
principle, implemented to brane states, opening the possibility
 for alternative dark matter candidates, such as the radion or the
 lightest odd KK in the presence of KK parity symmetry. For example, in~\cite{Anchordoqui:2023tln} we assume that no such
discrete symmetry survives and KK modes decay to the radion. In this
model we can avoid the velocity kick constraints on KK momentum
violation because the graviton tower decays into the radion before
primordial nucleosynthesis. It is of
interest to see whether it is possible to translate the KK momentum
violation effects into an effective $\delta_n$. We plan to elaborate on this idea in future publications.

\section{Conclusions}
\label{sec:7}

We have shown that 5D black holes perceiving the dark dimension are bigger, colder, and longer-lived
than usual 4D black holes of the same mass. Adopting the robust
Hawking's semiclassical approximation we have demonstrated that a
PBH all-dark-matter interpretation would be possible for the following
mass ranges:
\begin{itemize}
\item Schwarzschild black holes localized on the brane  $\Rightarrow 10^{15} < M_{\rm
    BH} /{\rm g} < 10^{21}$; 
\item  Schwarzschild black holes localized in the bulk $\Rightarrow 10^{11} \alt M_{\rm
    BH} /{\rm g} < 10^{21}$;
\item near-extremal black holes localized on the brane $\Rightarrow 10^{5} \sqrt{\beta} < M_{\rm
    BH} /{\rm g} < 10^{21}$;
\end{itemize}
where $\beta$ measures the near-extremality. The more speculative memory burden effect, with decay rate suppress by
$1/S_{\rm BH}$, could only extend this mass range if black holes are localized on
the bulk. Finally, we have argued that the mass ranges given above for
a PBH dark matter interpretation would not be modified if one assumes
a space with two boundaries for the dark dimension.

\section*{Acknowledgements}
We thank Gia Dvali, Elias Kiritsis and Georges Obied for discussion. The work of L.A.A. is supported by the U.S. National Science
Foundation (NSF Grant PHY-2112527). I.A. is supported by the Second
Century Fund (C2F), Chulalongkorn University. The work of D.L. is supported by the Origins
Excellence Cluster and by the German-Israel-Project (DIP) on Holography and the Swampland.

\section*{Appendix: Transition between 4D $\leftrightharpoons$ 5D black holes and related species description}

In this Appendix, we compare the differences in the change of the black
hole lifetime for the following two scenarios: {\it (i)}~addition of KK graviton
towers that inevitably emerge when the 4D spacetime is endowed with
compact
space and {\it (ii)}~a 4D effective theory with $N$ massless species
(but without reference to higher dimensions). 

We first recapitulate the properties of  higher dimensional black
holes. From
(\ref{rs}), the
Schwarzschild radius of a $d$ dimensional black hole scales as
$r_s \sim M_{\rm BH}^{1/(d-3)} /M_d^{(d-2)/(d-3)}$, where $M_d
\sim m_{\rm KK}^{(d-4)/(d-2)} M_p^{2/(d-2)}$ is the 5D Planck scale and $m_{\rm KK}$ is the KK mass scale.  The species scale can be rewritten in terms of  the number of KK species $N$ as
$M_d \sim M_p N^{-1/2}$
and we can also express $N$ in terms of the KK mass, yielding $N \sim
(M_p/m_{\rm KK})^{2(d-4)/(d-2)}$. Putting these formulas together $
r_s$ can be rewritten in terms of $N$ as
$
r_s \sim (M_{\rm BH}^{1/(d-3)}/ M_p^{(d-2)/(d-3)}) N^{(d-2)/(2d-6)}$
and we see that $r_s$ scales with a positive power of $N$. For example, for the dark
dimension scenario (with $d=5$), one gets
\begin{equation}
r_s = M_{\rm BH}^{1/2} \ M_p^{-3/2} \ N^{3/4} = M_{\rm BH} \ M_p^{-2}
(M_p/M_{\rm BH})^{1/2} \ N^{3/4} \,,
\end{equation}
where the term  $(M_{p}/M_{\rm BH})^{1/2} N^{3/4}$ is the enhancement
factor of the Schwarzschild radius compared to the 4D case. For the
dark dimension, $N \sim
10^{18}$ and  $(M_p/M_{\rm BH}) \sim 10^{-19}$ for $M_{\rm
  BH} \sim 10^{14}~{\rm g}$. Thus, we see that the radius is
increased compared to the 4D case.

Next, we calculate the Hawking decay time
$
\tau_{\rm H} \sim S_{\rm BH} r_s$,
where the entropy is
\begin{equation}
  S_{\rm BH} \sim M_{\rm BH} \ r_s \sim (M_{\rm BH}/M_d)^{(d-2)/(d-3)}
  \, .
\label{entropy_appen}  
\end{equation}
Thus, putting all this together we obtain
$
\tau_{\rm H} = (M_{\rm BH}^{(d-1)/(d-3)}/ M_d^{(2d-4)/(d-3)})
N^{(d-2)/(d-3)}$, 
which scales with a positive power of $N$. For example, for the dark
dimension (with $d=5$), it follows that
\begin{equation}
\tau_{\rm H} \sim M_{\rm BH}^2 \ M_{p}^{-3} \ N^{3/2} = M_{\rm BH}^3 \
M_{p}^{-4} \   (M_{p}/M_{\rm BH}) N^{3/2} \,,
\end{equation}
where the term  $(M_{p}/M_{\rm BH}) N^{3/2}$ is the enhancement factor
of the black hole lifetime compared to the 4D case.

Now, consider a 4D effective theory with $N$ species (and no reference to
higher dimensions). The minimal black hole radius is set by the
species length, $r_{\rm min} = 1/M_d = \sqrt{N} l_p$, with $l_p \sim
1/M_p$~\cite{Basile:2024dqq}. The corresponding minimal mass is $M_{\rm BH, min} = r_{s,{\rm min}}
M_p^2 = \sqrt{N} M_p$, and the corresponding entropy is $
S_{\rm BH, min} = r_{s, {\rm min}}^2 M_p^2 = N$.
Finally, the maximal temperature, which sets the semiclassical decay rate, is 
$ 
  T_{\rm H, max}= 1/r_{s,{\rm min}} = M_d = M_p/\sqrt{N}$. We can now
  calculate the lifetime of the minimal black hole
\begin{equation}
\tau_{\rm H, min} = r_{s,{\rm min}} \ S_{\rm BH, min} = N^{3/2} l_p \,,
\end{equation}
and see that it has increased with respect to the lifetime of the
minimal black hole we would have obtained in the absence of
$N$. However, it is important to stress that this enhancement is in
essence different from the one
associated to the KK towers, because
we are not comparing black holes of the same mass, i.e. the minimal
black hole in a 4D theory with species is heavier than the one in the
absence of $N$. For equal black
hole masses, the rate of decay in a theory with species is increased $N$-fold. As a consequence
the Hawking-decay time in the presence of species is shortened to~\cite{Alexandre:2024nuo}
\begin{equation}
  \tau_{\rm H} \sim r_s \ S_{\rm BH}/N \, .
\end{equation}  

\begin{figure}[htb!]
  \begin{minipage}[t]{0.49\textwidth}
    \postscript{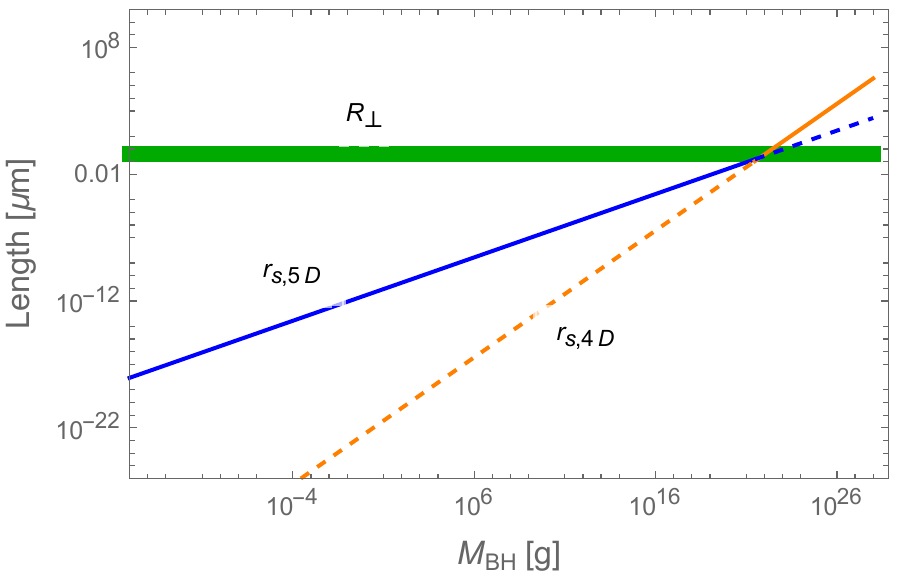}{0.9}
  \end{minipage}
\begin{minipage}[t]{0.483\textwidth}
    \postscript{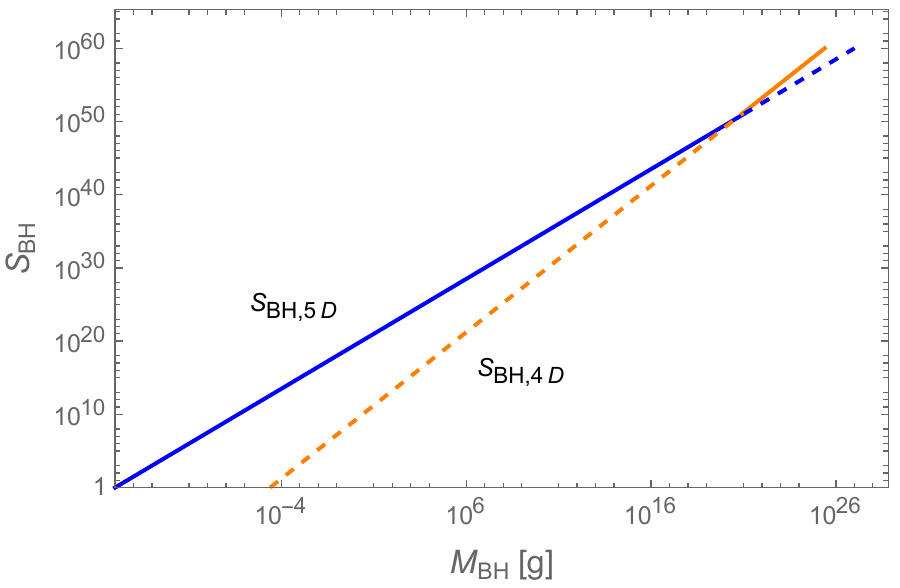}{0.9}
  \end{minipage}
  \caption{Scaling of the Schwarzschild radius (left) and black hole
    entropy (right) in $d=4$ and $d=5$
  dimensions. \label{fig:4}}
\end{figure}

To understand the difference between the two scenarios we can compare
(\ref{entropy_appen}) with the black hole entropy 
\begin{equation}
  S_{\rm BH} \sim (M_{\rm BH}/M_p)^2 \,,
\label{aa}
\end{equation}
  of a 4D effective theory with $N$
species but without compact dimensions. Setting $d=5$ and using $M_p^2 
\sim M_d^3 R_\perp$ we recast (\ref{entropy_appen}) as   
\begin{equation}
  S_{\rm BH} \sim M_{\rm BH}^{3/2} \ R_\perp^{1/2}/M_p \, .
\label{bb}
\end{equation}

In Fig.~\ref{fig:4} we show a comparison of the 4D and 5D scaling
behavior of the black hole entropy as given by (\ref{aa}) and (\ref{bb}). We can conclude that when considering a 4D theory with addition of $N$ species,
the black hole formulas do not change in essence, but there are more
channels for the evaporation process. For example, we can compare the
evaporation of a black hole into SM fields and its minimal extension
when the theory is extended by
three right-handed Dirac neutrinos. If
the black hole can emit right-handed neutrinos it would have a shorter
lifetime than if it only emits SM degrees of freedom. The minimum black
hole mass available in the SM extension is increased, because the black hole
must store extra information, this is all. Now, by adding species in a
higher dimensional theory, it follows from (\ref{aa}) and (\ref{bb})
that  the scaling behavior of
the entropy changes and for the black hole it is more convenient to be
in the 5D configuration because it has more entropy than the 4D
configuration. As can be seen in Fig.~\ref{fig:4}, for $R_\perp \sim 1~\mu{\rm m}$, the transition takes
place at $M_{\rm BH} \sim 10^{21}~{\rm g}$.  Note that the two
entropies in (\ref{aa}) and (\ref{bb}) are equal at the 5D-4D transition point
where $M_{\rm BH} = M_p^2 R_\perp$. Moreover, the entropy crosses the
horizontal axis 
where the black hole masses are the same as the 4D or 5D Planck masses. Then, the associated lengths are the
4D or 5D Planck lengths, where the two entropies are equal to one.\footnote{The phase
  transition between 4D and 5D black holes was also recently
  investigated in~\cite{Bedroya:2024uva}.}

\end{document}